\documentclass[11pt]{article}
\usepackage{graphicx}
\usepackage{dcolumn}
\usepackage{amsmath}

\newcommand{\BABARPubYear}    {02}

\newcommand{\BABARConfNumber} {014}
\newcommand{\SLACPubNumber} {9299}

\usepackage{epsfig}
\usepackage{feynmp}

\input pubboard/babarsym

\def\Dstarp     {\ensuremath{D^{*+}}\xspace}
\def\Dstarm     {\ensuremath{D^{*-}}\xspace}

\def\Dp         {\ensuremath{D^+}\xspace}
\def\Dm         {\ensuremath{D^-}\xspace}

\def\DeltaEStd  {\ensuremath{\Delta E} \xspace}

\def\masslik    {\ensuremath{{\cal L}_{Mass}}\xspace}

\def\leptontag{{\tt Lepton}}
\def\kaonitag{{\tt Kaon\,I}}
\def\kaoniitag{{\tt Kaon\,II}}
\def\othertag{{\tt Inclusive}}

\long\def\inst#1{\par\nobreak\kern 4pt\nobreak
    {\it #1}\par\vskip 10pt plus 3pt minus 3pt}

\begin{document}
{\pagestyle{empty}

\begin{flushright}
\babar-CONF-\BABARPubYear/\BABARConfNumber \\
SLAC-PUB-\SLACPubNumber \\
July 2002 \\
\end{flushright}

\par\vskip 5cm

\begin{center}
\Large \bf Measurement of Time-Dependent \CP Asymmetries\\
and the \CP-odd Fraction
in the Decay \Bztodstdst
\end{center}
\bigskip

\begin{center}
\large The \babar\ Collaboration\\
\mbox{ }\\
July 24, 2002
\end{center}
\bigskip \bigskip

\begin{center}
\large \bf Abstract
\end{center}
We present a measurement of time-dependent $CP$
asymmetries and an updated determination of the $CP$-odd fraction in the decay
$B^0 \rightarrow D^{*+}D^{*-}$.
The measurements are derived from a data sample 
of $88 \times 10^{6} B\bar{B}$ pairs collected 
by the {\mbox{\slshape B\kern-0.1em{\smaller A}\kern-0.1em
    B\kern-0.1em{\smaller A\kern-0.2em R}}} detector at the
PEP-II energy asymmetric $B$ Factory at SLAC.
All results are preliminary.

\vfill
\begin{center}
Contributed to the 31$^{st}$ International Conference on High Energy Physics,\\ 
7/24---7/31/2002, Amsterdam, The Netherlands
\end{center}

\vspace{1.0cm}
\begin{center}
{\em Stanford Linear Accelerator Center, Stanford University, 
Stanford, CA 94309} \\ \vspace{0.1cm}\hrule\vspace{0.1cm}
Work supported in part by Department of Energy contract DE-AC03-76SF00515.
\end{center}

\newpage
} 

\begin{center}
\small

The \babar\ Collaboration,
\bigskip

B.~Aubert,
D.~Boutigny,
J.-M.~Gaillard,
A.~Hicheur,
Y.~Karyotakis,
J.~P.~Lees,
P.~Robbe,
V.~Tisserand,
A.~Zghiche
\inst{Laboratoire de Physique des Particules, F-74941 Annecy-le-Vieux, France }
A.~Palano,
A.~Pompili
\inst{Universit\`a di Bari, Dipartimento di Fisica and INFN, I-70126 Bari, Italy }
% ...

\end{center}\newpage

\section{Introduction}
\label{sec:Introduction}
The symmetry for combined charge conjugation {\it (C)} and parity {\it (P)} transformations 
is violated in $B$ decays. Measurements of \CP asymmetries by the
\babar~\cite{babarCP} and BELLE~\cite{belleCP} collaborations
established this effect and are compatible with
the Standard Model expectation based on the current knowledge of 
the Cabibbo-Kobayashi-Maskawa~\cite{CKM} quark-mixing matrix elements.

As a result of the interference between direct $B$ decay, expected to be dominated
by the
tree decay diagram, and decay after flavor change, a \CP -violating  asymmetry is expected in the
time evolution of the decays\footnote{Charge-conjugate modes are implied throughout this paper.}
$\Bz \to D^{*+} D^{*-}$, within the framework of the Standard Model~\cite{aleksan}.
Up to corrections due to theoretically uncertain penguin diagram
contributions~\cite{sanda}, this \CP asymmetry is related 
to \stwob
($\beta \equiv \arg \left[\,
  -V_{\rm cd}^{}V_{\rm cb}^* / V_{\rm td}^{}V_{\rm tb}^*\, \right]$).
Penguin-induced corrections are predicted to be small
in models based on the factorization approximation 
and heavy quark symmetry; an effect of about $2\%$ is predicted by Ref.~\cite{xing}. 
A comparison of measurements of \stwob~\cite{babar-stwob-newprl}
from charmonium-containing
$b \to c \bar{c} s$ modes such as $B^0 \to J/\psi K^0_S$, with that 
obtained in \Bztodstdst is an important test of these models and the 
consistency of the Standard Model.

The \Bztodstdst mode is a pseudoscalar decay to a vector-vector final state,
with contributions from three partial waves with different \CP
parities: even for the $S$- and $D$-waves, odd for the $P$-wave.
In the model described in Ref.~\cite{xing2} the $P$-wave contribution
is predicted to be about $11\%$. 
The angular distribution of the decay products
can be used to 
measure the \CP parameters of 
the \CP-even and \CP-odd components
\cite{angular}.

Following our initial results on this channel~\cite{ref:dstdstprl}, 
we present here an updated determination
of the \CP -odd fraction $R_\perp$ in the decay \Bztodstdst in \babar,
based on a projected one-dimensional time-integrated angular analysis.  
We also present a preliminary measurement of 
the time-dependent \CP asymmetry, obtained from 
a combined analysis of the time dependence 
of flavor-tagged decays
and the one-dimensional angular distribution of decay products.

\section{The \babar\ Detector and Dataset}
\label{sec:babar}
The data used in this analysis were collected with the \babar\ detector
at the \pep2 storage ring.  The data sample used for the
time-dependent \CP-asymmetry measurement corresponds to 
$88.0 \times 10^6 $
$e^+e^- \to \Upsilon(4S) \to B\bar{B}$ events 
and the sample used for the $R_\perp$
measurement corresponds to 
$84.4 \times 10^6 B\bar{B}$ pairs.
The collider is operated with
asymmetric beam energies, producing a boost $(\beta\gamma = 0.55)$ of
the \FourS along the collision axis. 

\babar\ is a solenoidal detector optimized for the asymmetric beam
configuration at \pep2, and is described in detail
elsewhere~\cite{ref:babar}.  Charged particle (track) momenta are
measured in a tracking system consisting of a 5-layer, double-sided,
silicon vertex tracker (SVT) and a 40-layer drift chamber (DCH) filled
with a gas 
mixture of helium and isobutane, both operating within a 1.5\,T
superconducting solenoidal magnet.  Photon candidates are selected as
local maxima of deposited energy in an electromagnetic calorimeter
(EMC) consisting of 6580 CsI(Tl) crystals arranged in barrel and
forward endcap subdetectors.  In this analysis, tracks are identified
as pions or kaons by the Cherenkov angle $\theta_c$, measured using a
detector of internally reflected Cherenkov light (DIRC), and by the
energy deposition, \dedx, in the tracking system.  The flux
return of the magnet is instrumented with resistive plate chambers
interspersed with iron (IFR) for the identification of muons and
long-lived neutral hadrons.

\section{Event Selection}
\label{sec:EventSelection}
\Bz mesons are exclusively reconstructed by combining two charged $D^{*}$
candidates reconstructed in a number of \Dstar and $D$ decay modes.
Events are pre-selected by requiring that the normalized second Fox-Wolfram
moment~\cite{ref:fox} of the event be less than 0.6.
We also require
that the cosine of the angle between the thrust axis of the reconstructed $B$
and the thrust axis of the rest of the event be less than 0.9.
These criteria are used to reject events coming from the $\epem \to
\ccbar$ continuum process. Backgrounds from $u, d, s$ continuum
processes are negligible in this analysis due to the presence of two
charm particles in the final state.

The \Dz and \Dp modes reconstructed are $\Dz \to \Km \pip$, $\Dz \to \Km \pip
\piz$, $\Dz \to \Km \pip \pip \pim$, $\Dz \to \KS \pip \pim$, $\Dp \to
\Km \pip \pip$, $\Dp \to \KS \pip$ and $\Dp \to \Km \Kp \pip$.

Charged kaon candidates are required to be inconsistent with the pion
hypothesis, as inferred from the Cherenkov angle
measured by the DIRC and the specific ionization measured by the SVT and DCH.  
No particle identification requirements are made
for the kaon from the decay $\Dz \to \Km \pip$.

$\KS \to \pip\pim$ candidates are required to have an invariant mass
within 25\,\mevcc of the nominal \KS mass~\cite{pdg}.  The angle between the
flight direction and the momentum vector of the \KS candidate is
required to be less than 200\,\mrad, and the transverse flight distance
from the primary event vertex must be greater than 2\,mm.
A mass-constrained fit is then applied to each surviving  $\KS$ candidate, in order
to improve the mass resolution of the $\KS \pip$ and $\KS \pip\pim$ combinations.

Neutral pion candidates are formed from
two photons in the EMC, each with energy above 30\,\mev;
the invariant mass of the pair must be within 20\,\mevcc\ of the nominal $\pi^0$ mass,
and their summed energy must be greater than 200\,\mev.
A mass-constrained fit is then applied to these $\pi^0$ candidates. 
The $\pi^0$ from $D^{*+} \to D^+ \piz$ decay (``soft'' $\pi^0$),
however, is required to have an invariant mass within
35\,\mevcc\ of the nominal $\pi^0$ mass and momentum in the \FourS frame
in the interval $70 < |p^*| < 450\mevc$, with no requirement on the photon energy sum.

\Dz and \Dp meson candidates are required to
have an invariant mass within 20\,\mevcc of the nominal \Dz or \Dp mass.
The same interval is used for all \Dz modes except $\Km\pip\piz$, 
which has a looser requirement of 35 \mevcc\ due to the momentum resolution
of the $\piz$.

\Dstarp mesons
are reconstructed in their decays $\Dstarp\to\Dz\pip$ and $\Dstarp\to\Dp\piz$. 
We include $\Dstarp\Dstarm$ combinations decaying 
to $(\Dz\pip, \Dzb\pim)$ or $(\Dz\pip, \Dm\piz)$, but not $(\Dp\piz,\Dm\piz)$ due to 
the smaller branching fraction and larger expected backgrounds.
The \Dz and \Dp candidates are subjected to a mass-constrained fit and
then combined with soft pion candidates.  A vertex fit is performed
that includes the position of the beam spot to improve the angular
resolution of the soft pion.

If an event contains both
a \Dstarp and a \Dstarm candidate, each is subjected to a mass
constraint fit, and then combined to form a $B$ candidate.
 \Bz candidates with well reconstructed \Dstar and $D$
mesons are selected by means of a mass likelihood variable, 
\masslik, that includes all
measured \Dstar and $D$ masses.
This variable is defined in Eq.~\ref{eq:masslikelihood}: 
   
\begin{eqnarray}
\label{eq:masslikelihood}
\masslik & = &G(m_D ; ~m_{D_{PDG}}, \sigma_{m_D}) \times
        G(m_{\Db} ; ~ m_{D_{PDG}}, \sigma_{m_{\Db}}) \times \nonumber\\
        & & \left[ f_{core} ~ G(\Delta m_{\Dstarp} ; ~\Delta m_{\Dstar_{PDG}}, 
             \sigma_{\Delta m_{core}})  \right. \nonumber \\
&&\qquad \left. + \quad (1-f_{core}) ~  G(\Delta m_{\Dstarp} ; ~\Delta m_{\Dstar_{PDG}}, 
             \sigma_{\Delta m_{tail}}) \right] \times \nonumber\\
        & & \left[ f_{core} ~ G(\Delta m_{\Dstarm} ; ~\Delta m_{\Dstar_{PDG}}, 
             \sigma_{\Delta m_{core}})  \right. \nonumber \\
&&\qquad \left. + \quad (1-f_{core}) ~ G(\Delta m_{\Dstarm} ; ~\Delta m_{\Dstar_{PDG}}, 
             \sigma_{\Delta m_{tail}}) \right] 
\end{eqnarray}
where $G(x;\mu,\sigma)$ is a normalized Gaussian defined with mean $= \mu$ and
RMS $= \sigma$; the subscript PDG refers to the nominal value~\cite{pdg}.
For $\sigma_{m_D}$ we use errors calculated candidate-by-candidate. 
The parameter $f_{core}$ is the ratio of areas for the 
core and tail Gaussians.  
This along with $\sigma_{\Delta m_{core}}$ and $\sigma_{\Delta m_{tail}}$
are determined from fitting the $\Delta m$ distributions in simulated
signal events.
The value of $-\ln(\masslik)$ is used to select signal in preference
to background, with a different requirement used for each $D$ decay mode
combination.
In an event where more than one $B$ candidate is reconstructed,
the candidate with the lowest $-\ln(\masslik)$ value is chosen.

The primary variables used to distinguish signal from background are
the energy-substituted mass,
$$\mes \equiv \sqrt{{E_{Beam}^*}^2 - {p_B^*}^2}$$
and the difference of the $B$ candidate energy from beam energy,
$$\DeltaEStd \equiv E_{B}^* - E_{Beam}^* $$
where the star indicates variables evaluated in the \FourS center-of-mass frame. 
The fits described in the following sections
are performed on events required to have $|\DeltaEStd| < 25 \mev$ and
$\mes > 5.2 \gevcc$ .
The requirements on $-\ln(\masslik)$ and \DeltaEStd were chosen to optimize
$S^2/(S+B)$, where $S$ is the expected number 
of signal events and $B$ is the expected number of background events.
The optimization process was done with
samples of simulated signal events and with generic \BB and \ccbar Monte Carlo simulated events.  

\begin{figure}[!ht]
\begin{center}
\epsfig{file=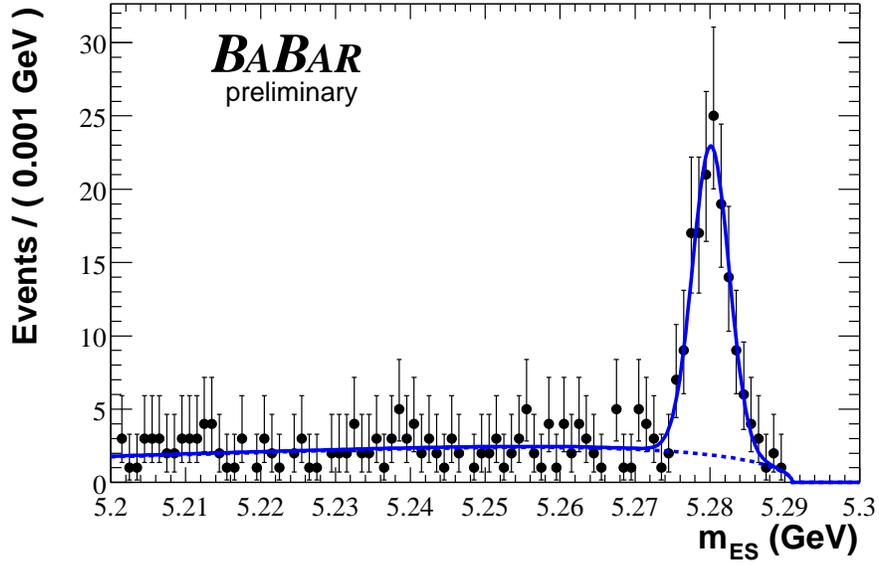,width=12cm}
\caption{The $m_{ES}$ projection of the data for $\Bztodstdst$.  These events
are required to have $|\DeltaEStd| < 25 \mev$.  The solid (dashed) line represents
the result (background component) of the fit described in the text.
}
\label{fig:dstdst_mes}
\end{center}
\end{figure}

Figure~\ref{fig:dstdst_mes} shows the events in the
\mes projection after all selection criteria have been applied.
The fit to this distribution uses a Gaussian and an ARGUS
function~\cite{argus} parameterization for signal and background, respectively.
For the data sample corresponding to $84.4 \times10^6 B\bar{B}$ pairs, 
the fitted signal yield is $126 \pm 13$ $ \Dstarp\Dstarm$ events.

\section{Measurement of the \CP-odd Fraction in \Bztodstdst}
\label{sec:TransversityAnalysis}
\begin{figure}[!h]
\begin{center}
\epsfig{file=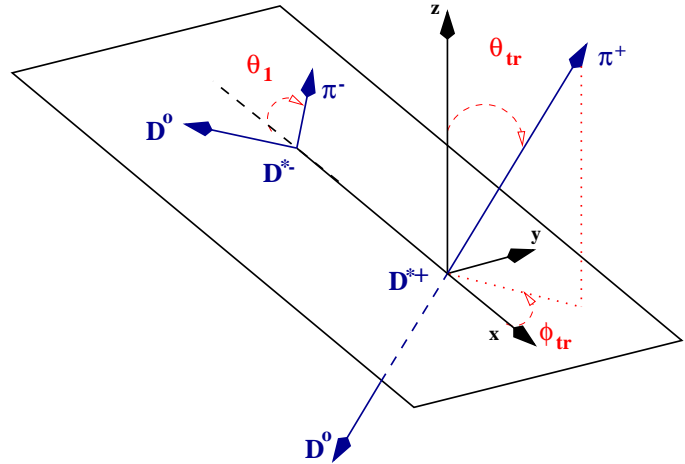,width=9cm}
\caption{Representation of the ``Transversity Frame'' for the decay \Bztodstdst\ .
The momenta of the $D^{*-}$ decay products
are represented in the $B^0$ rest frame,
while the momenta of the $D^{*+}$ decay products
are represented in the $D^{*+}$ rest frame.
The $x$ direction is defined by the direction of flight of the $D^{*+}$
in the $B^0$ rest frame.
The $(x,y)$ plane is defined by the momenta of the $D^{*-}$ decay products
in the $B^0$ rest frame.
The ``transversity'' axis, $z$, is orthogonal to the $(x,y)$ plane.
}
\label{fig:transframe_DD}
\end{center}
\end{figure}

 In this section we present a one-dimensional angular
analysis to determine 
the fraction, $R_\perp$, of the $P$-wave \CP -odd component
in the vector-vector final state of the \Bztodstdst decay.  
In the transversity basis~\cite{angular} the following three
angles involving decay products are defined (see Fig.~\ref{fig:transframe_DD}):
\begin{itemize}
\item
the polar angle $\theta_1$ between the momentum of the  
$\pi^-$ in the $D^{*-}$ rest frame,
and the direction of flight of the $D^{*-}$ in the $B$ rest frame,
\item
the polar 
angle $\theta_{\rm tr}$ between the normal, $z$, 
to the $D^{*-}$ decay plane and the 
$\pi^+$ line of flight in the $D^{*+}$ rest frame, and 
\item
the corresponding azimuthal angle $\phi_{\rm tr}$. 
\end{itemize}

\clearpage
The time-dependent angular distribution of decay products in the transversity
frame for the mode $B^0 \to D^{*+}D^{*-}$ 
is given by~\cite{penguin}\footnote{Eq.5.44 in the reference appears
  with wrong
  signs in the last two terms and has been corrected in
  Eq.~\ref{eq:angdist} of this paper.}:
\begin{eqnarray}
\displaystyle
\frac{1}{\Gamma} \frac{{\rm{d}}^4\Gamma}{{\rm{d}}\cos \theta_1 {\rm{d}}\cos \theta_{\rm tr} {\rm{d}} \phi_{\rm tr}{\rm{d}t}}  & = &\frac{9}{32\pi} 
\frac {1}{|A_0|^2 + |A_{\parallel}|^2 + |A_{\bot}|^2} \nonumber \\[2mm] 
&&\{\rule{0mm}{5mm} 
4 |A_0|^2 \cos^2 \theta_1 \sin^2 \theta_{\rm tr} \cos^2 \phi_{\rm tr}   \nonumber \\[2mm] 
&&+ 2 |A_{\parallel}|^2 \sin^2 \theta_1 \sin^2 \theta_{\rm tr} \sin^2 \phi_{\rm tr}  \nonumber \\[2mm]
&&+ 2 |A_{\bot}|^2 \sin^2 \theta_1 \cos^2 \theta_{\rm tr}  \nonumber \\[2mm]
&&+ \sqrt{2} {\rm Re} (A^*_{\parallel}A_{0}) \sin 2\theta_1 \sin^2\theta_{\rm tr} \sin 2\phi_{\rm tr}  \nonumber \\[2mm]
&&- \sqrt{2} {\rm Im} (A^*_0A_{\bot}) \sin 2\theta_1 \sin 2\theta_{\rm tr} \cos \phi_{\rm tr}  \nonumber \\[2mm]
&& - 2 {\rm Im} (A^*_{\parallel} A_{\bot}) \sin^2 \theta_1 \sin 2\theta_{\rm tr} \sin \phi_{\rm tr} \rule{0mm}{5mm} \}\ .  
\label{eq:angdist}
\end{eqnarray}
where $A_0, A_{\parallel}, A_{\bot}$ are the time-dependent decay amplitudes in the 
transversity basis.
For the \Bzb\ decay $\bar{A}_{\bot} = -A_{\bot}$, under the assumption
there is no direct \CP violation. 

In principle, 11 unknown parameters
can be extracted from a full angular, time, and $B$-flavor dependent analysis.
These parameters are:
$M_i \equiv |A_i(t=0)|$, 2  relative phases,
and the complex \CP parameters $\lambda_i = (q/p)(\bar{A_{i}}/A_{i})$,
with $i = \parallel,0,\perp$, resulting from the interference
of mixing ($q/p$) and decay amplitudes ($A_{i}$).
Considering the size of the current data sample,  
a simplified
strategy is adopted to extract only the \CP-odd component from 
the $\cos \theta_{tr}$ distribution.

Ignoring the $B$ flavor and integrating over time, $\theta_1$,
 and $\phi_{tr}$ results in the one-dimensional differential
decay rate:
  \begin{eqnarray}
& & {1 \over \Gamma} \ {{\rm{d}}\Gamma \over {\rm{d}}\cos \theta_{\rm tr}} = 
 {3 \over 4} (1-R_\perp) \sin^2 \theta_{\rm tr} + {3 \over 2} R_\perp \cos^2 \theta_{\rm tr}  
   \label{AngDisArt}
  \end{eqnarray}
with:
\begin{eqnarray}
R_\perp & = & {M_\perp^2 \over M_0^2 + M_\parallel^2 + M_\perp^2}\ .
 \end{eqnarray}

The measurement of $R_\perp$ is based on an
unbinned maximum likelihood fit of the $\cos\theta_{ \rm tr}$ distribution, with a simultaneous fit to the
$m_{ES}$ distribution.  The probability density function for the $m_{ES}$ distribution is given 
by the sum of normalized ARGUS~\cite{argus} and Gaussian functions; the relative weight of each function
is given by a signal fraction $f_{sig}$, which is allowed to float in the likelihood fit.  
The likelihood is defined as:
\begin{equation}
\begin{array}{r@{}c@{}l}
\displaystyle
{\cal{L}}=
\prod_{i=1,n}
\left[f_{sig}\rule{0mm}{4mm}\right.
&\times&
\left.   {\cal{F}}_{\mes}^{sig}(\mes_{, i};\sigma_{m_{ES}}) \times  {\cal{F}}_{R_\perp}(\cos\theta_{i};R_\perp)+ \right.\\[5mm]
(1-f_{sig})
&\times&
\left. {\cal{F}}_{\mes}^{bkg}(\mes_{, i};\kappa) \times  {\cal{F}}_{bkg}(\cos\theta_{i};b_2)
\rule{0mm}{4mm}\right],
\end{array}
\label{ll1}
\end{equation}
where $n$ is the number of selected events in the \mes distribution,
${\cal{F}}_{\mes}^{sig} $ is the signal Gaussian for the \mes 
distribution and 
$ {\cal{F}}_{\mes}^{bkg}$ is the background ARGUS function with shape parameter 
$\kappa$.
${\cal{F}}_{R_\perp} $ refers to the probability density function for signal
events, given by Eq.~\ref{AngDisArt}.
The background
shape is modeled by a polynomial in $\cos\theta_{\rm tr}$:
\begin{equation}
\displaystyle
{\cal{F}}_{bkg}(\cos\theta_{tr};b_2) = N \times (1 + b_2 \cos^2\theta_{tr}),
\label{bkgPDF}
\end{equation}
where $N$ is the normalization factor.

We categorize our events in three types:
$\Dstarp\Dstarm \to (\Dz\pip,\Dzb\pim)$,  
$(\Dz\pip,\Dm\piz)$, and $(\Dp\piz,\Dzb\pim)$.
We distinguish these three types of events because events with a
neutral slow pion and events with a charged slow pion have different
background levels and $\cos\theta_{\rm tr}$ resolution (note: $\theta_{\rm tr}$ is
the angle between the slow pion from the $D^{*+}$ and the transversity
axis).

We allow different signal fractions for each event type.
Thus, the parameters floating in the likelihood fit are: 3 signal
fractions, the background 
parameter $b_2$, 
3 \mes parameters ($\sigma$ and mean of Gaussian fit, and $\kappa$ from ARGUS
shape), and $R_\perp$.

\begin{figure}[!ht]
\begin{center}
\epsfig{file=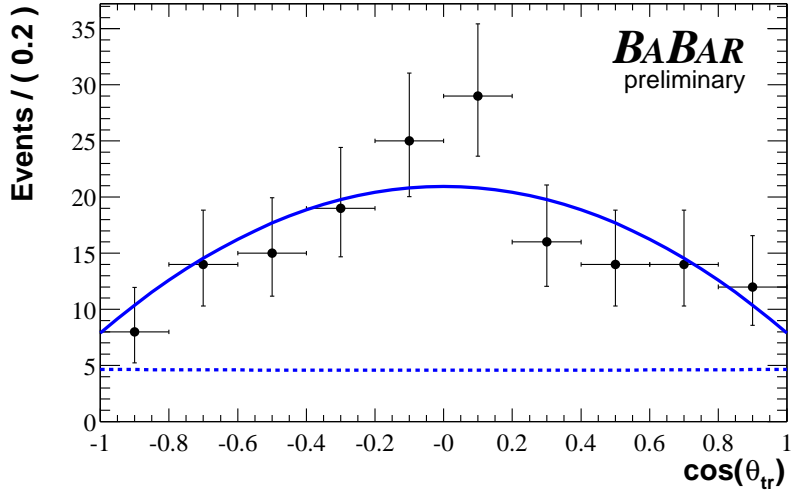,width=11cm}
\caption{ Likelihood fit result to the $\cos\theta_{\rm tr}$ distribution of the 
$D^{*+}D^{*-}$ events. The data points
shown are from the region $\mes > 5.27\,\gevcc$ and the solid line
is the projection of the fit result in the same region.  The dotted line
represents the component of the pdf for the background.
}
\label{fig:datafit}
\end{center}
\end{figure}

A fit to the dataset yields a 
value of $R_\perp = 0.096 \pm 0.060(stat)$, neglecting possible biases
from angular resolution in $\theta_{tr}$ and detector acceptance.
Figure~\ref{fig:datafit} shows the distribution of $\cos\theta_{\rm tr}$ for
candidates with \mes  in the 
range $\mes > 5.27\,\gevcc$, with the result of the fit projected
in the same region.
The background component of 
the pdf is shown as the dotted line.   

The experimental resolution of $\theta_{\rm tr}$ biases the measurement of $R_\perp$; 
the simulated distribution of residuals on $\theta_{\rm tr}$ 
is seen to have significant tails
caused by mis-reconstructed events.  The presence of these tails
distorts the $\cos\theta_{\rm tr}$ distribution, and thus produces a bias on
$R_\perp$.  
A smaller bias on $R_\perp$ is produced by the
transverse momentum dependence of the detector reconstruction efficiency.
The presence of the two slow pions from the \Dstar decays
makes this analysis susceptible to this effect.
The net bias from these two effects was estimated from
a study on simulated data, and is found to be +0.028.  This was determined using
simulated events with a generated value of $R_\perp$ very similar to the 
value measured in data.   The corresponding systematic error incurred
for this correction is taken conservatively to be the full size of the
correction, namely 0.028.  The next largest systematic uncertainty 
affecting the $R_\perp$ measurement arises from
the background parameterization (0.005).  The total 
systematic uncertainty on $R_\perp$ is determined to be 0.03, giving as a preliminary result the final corrected
value:
\begin{equation}
\displaystyle
R_\perp = 0.07 \pm 0.06(stat) \pm 0.03(syst). 
\end{equation}
This is an update to a previous \babar\
measurement~\cite{ref:dstdstprl}, with a factor of three reduction in the statistical
error ($R_\perp$ dependent).

\section{Measurement of the Time-Dependent \CP Asymmetry in \Bztodstdst}
\label{sec:CPasymmetryAnalysis}
In this section we present a determination of the time-dependent \CP asymmetry
in \Bztodstdst. We perform
a combined analysis of the $\cos \theta_{\rm tr}$ distribution of 
all selected events and 
the time dependence of those with a flavor tag.
The inclusion of the angular dependence enables us to fit 
for the \CP asymmetries of the \CP-even and \CP-odd components separately.

Although factorization models predict a small
penguin contamination of the weak phase difference ${\rm Im}(\lambda_{f})=-\sin2\beta$~\cite{xing},
a sizeable penguin diagram contribution cannot {\it a priori} be excluded.
Thus, the value of 
$\lambda_{f}=\frac{q}{p}\frac{\bar{A}(f)}{A(f)}$ 
can be different for the three transversity amplitudes because of
possible different
penguin-to-tree ratios. These contributions are explicitly included in the 
parameterization of the flavor-tagged decay rates described here.

The decay rate, $f_+ (f_-)$,  for a neutral $B$ meson tagged as a $B^{0}
(\bar{B}^0)$ 
can be obtained from Eq.~\ref{eq:angdist} as:
\begin{eqnarray}
f_\pm(\deltat) & = &
\frac{{\rm e}^{ - | \deltat |/\tau_{B^0} }}{4\tau_{B^0}}
 \Bigl\{ O(1-{\textstyle{\frac 12}}\Delta {\cal D}) \mp   {\cal D} \left[
 S \sin{ (\deltamd  \deltat) } 
+ C \cos{ (\deltamd  \deltat) }  \right]  \Bigr\},
 \label{eq:sincos}
\end{eqnarray}

\vskip12pt\noindent
where $\Delta t = t_{\rm rec} - t_{\rm tag}$ is the difference between 
the proper decay time of the reconstructed $B$ meson ($B_{\rm rec}$) and 
the proper decay time of the tagging $B$ meson ($B_{\rm tag}$),
$\tau_{\Bz}$ is the \Bz lifetime, and \deltamd 
is the mass difference determined from the \Bz-\Bzb oscillation frequency.
The dilution factor, ${\cal D}$, describes the effect of incorrect tags,
with $\Delta {\cal D}$ accounting for possible differences in the mis-tag
probability between $\Bz$ and $\Bzb$.  
The $O$, $C$ and $S$ coefficients are defined as:

\begin{eqnarray} 
O &=& \frac{3}{4} [(1-R_\perp) \sin^2\theta_{\rm tr} 
   +    2  R_\perp \cos^2\theta_{\rm tr}]
  \nonumber \\
C &=& \frac{3}{4} [(1-R_\perp)\frac{1 - |\lambda_{+}|^{2}}{1 +
  |\lambda_{+}|^{2}}  \sin^2\theta_{\rm tr}  +
 2 R_\perp \frac{1 - |\lambda_{\bot}|^{2}}{1 + |\lambda_{\bot}|^{2}}  \cos^2\theta_{\rm tr}] \nonumber \\
 S &=& -\frac{3}{4}[(1-R_\perp)  \frac{2{\rm Im}(\lambda_{+})}{1 + |\lambda_{+}|^{2}} \sin^2\theta_{\rm tr} -
 2 R_\perp \frac{2{\rm Im}(\lambda_{\bot})}{1 +  |\lambda_{\perp}|^{2}} \cos^2\theta_{\rm tr} ].
\label{eq:ocs}
\end{eqnarray}
These coefficients
contain the explicit dependence on the transversity angle $\theta_{\rm tr}$
defined in the previous section, which provides a separation
between the \CP -odd ($\cos^2\theta_{\rm tr}$)
and \CP -even  ($\sin^2\theta_{\rm tr}$) components.

The \CP-even parameters, $|\lambda_{+}|$ and ${\rm Im}(\lambda_{+})$, 
are related to $\lambda_{\parallel}$ and $\lambda_{0}$ by:

\begin{eqnarray}
 \frac{1-|\lambda_{+}|^2}{1+|\lambda_{+}|^2}&=&
 \frac{
 \frac{1-|\lambda_{\parallel}|^2}{1+|\lambda_{\parallel}|^2}
 M_{\parallel}^2+ 
 \frac{1-|\lambda_{0}|^2}{1+|\lambda_{0}|^2} M_{0}^2 }
      {M_{\parallel}^2 +M_{0}^2} \nonumber \\ 
 \frac{{\rm Im}(\lambda_{+})}{1+|\lambda_{+}|^2} &=&
 \frac{
 \frac{{\rm Im}(\lambda_{\parallel})}{1+|\lambda_{\parallel}|^2}M_{\parallel}^2+\frac{{\rm Im}(\lambda_{0})}{1+|\lambda_{0}|^2}M_{0}^2}{M_{\parallel}^2 +M_{0}^2}. 
\end{eqnarray}
It should be noted that this formulation
does not take into account detector acceptance.  
Therefore, a combined fit to the  $\Delta t$ and $\cos\theta_{\rm tr}$ 
dependence of the data will give an ``effective'' 
value of $R_\perp$, which is not necessarily identical to the 
acceptance-corrected value from the time-integrated measurement.

A measurement of \CP asymmetries requires a determination of the experimental
$\Delta t$ resolution and the fraction of events in which the tag
assignment is incorrect. A mis-tag fraction $w$ reduces the observed
\CP asymmetry by a factor ${\cal D} = 1-2w$.
The mis-tag fractions and $\Delta t$ resolution functions are
determined from a sample, $B_{\rm flav}$, of neutral $B$ decays to
flavor eigenstates ($D^{(*)-}h^+ (h^+=\pi^+,\rho^+$, and $a_1^+)$ and $\jpsi\Kstarz
(\Kstarz\to\Kp\pim)$)  as for the \stwob measurement using
charmonium decays, described in detail
elsewhere~\cite{babar-stwob-newprl}.

\par
\begin{table}[!t]
\caption
{Efficiencies $\epsilon_i$, average mis-tag fractions $\mistag_i$, mis-tag fraction differences
$\Delta\mistag_i=\mistag_i(\Bz)-\mistag_i(\Bzb)$, and $Q$ extracted for each tagging
category $i$ by using the $B_{\rm flav}$ 
sample. 
}
\label{tab:mistag}
 \begin{center}
\begin{tabular}{|l|r|r|r|r|}
\hline
Category     & 
$\ \ \ \varepsilon$   (\%) & 
$\ \ \ \mistag$       (\%) & 
$\ \ \ \Delta\mistag$ (\%) &
$\ \ \ Q$             (\%) \\
\hline\hline
\leptontag   & $ 9.1\pm0.2$ & $ 3.3\pm 0.6 $ & $-1.5\pm1.1$ & $ 7.9\pm0.3$ \\  
\kaonitag    & $16.7\pm0.2$ & $10.0\pm 0.7 $ & $-1.3\pm1.1$ & $10.7\pm0.4$ \\ 
\kaoniitag   & $19.8\pm0.3$ & $20.9\pm 0.8 $ & $-4.4\pm1.2$ & $ 6.7\pm0.4$ \\ 
\othertag    & $20.0\pm0.3$ & $31.5\pm 0.9 $ & $-2.4\pm1.3$ & $
2.7\pm0.3$ \\
  \hline\hline
All          & $65.6\pm0.5$ &                &              &
$28.1\pm0.7$ \\ 
\hline
\end{tabular} 
\end{center} 
\end{table} 
We use multivariate algorithms to identify signatures of $B$ decays that
determine the flavor of $B_{\rm tag}$.
Primary leptons from semileptonic $B$ decays are selected
from identified electrons, muons, and isolated energetic tracks.
We use the charges of the best kaon candidates to define a kaon tag.
Soft pions from \Dstarp decays are selected on the basis
of their momentum and direction with respect to
the thrust axis of $B_{\rm tag}$.
A neural network, which combines the outputs of these algorithms,
takes into account correlations between different sources of
flavor information and provides an estimate of the mis-tag probability for each event.
\par
Using the outputs of the algorithms and the estimated mis-tag probability,
each event is assigned to one of four hierarchical, mutually exclusive tagging categories.
The \leptontag\ category contains events with an identified lepton, and a supporting
kaon tag if present.
Events with a kaon candidate and soft pion with opposite charge and
similar flight direction are assigned to the \kaonitag\ category.
Events with only a kaon tag are assigned to the \kaonitag\ or \kaoniitag\ 
category depending on the estimated mis-tag probability.
The \kaoniitag\ category also contains the remaining events with a soft pion.
All other events are assigned to the \othertag\
category or excluded from further analysis based
on the estimated mis-tag probability.
The tagging efficiencies $\eps_i$
for the four tagging categories are measured from data and summarized in·
Table~\ref{tab:mistag}.
The figure of merit for tagging is the effective tagging efficiency
$Q \equiv \sum_i {\eps_i (1-2\mistag_i)^2} $.
This algorithm improves $Q$ by about 7\% (relative) over the algorithm used in
Ref.~\cite{babar-stwob-prd}.

The algorithm for vertex reconstruction and the
determination of \deltat are described in detail in
Ref.~\cite{babar-stwob-prd}.     
The time interval \deltat between the two $B$ decays is calculated
from the measured separation \deltaz between the decay vertex of the 
reconstructed  $B$ meson and the vertex of the
flavor-tagging $B$ meson along the collision axis. 
We determine the $z$ position of the $B_{\rm rec}$ vertex from
the charged tracks that constitute the $B_{\rm rec}$ candidate. The
decay vertex of the $B_{\rm tag}$ is determined by fitting the tracks not
belonging to the $B_{\rm rec}$ candidate to a 
common vertex. 
An additional constraint on the tagging vertex comes from a pseudotrack 
computed from the  $B_{\rm rec}$ vertex and three-momentum,
the beam-spot 
and the \FourS
momentum. 
Events with a \deltat\ error of less than 2.5\ps, and a measured $\vert \deltat \vert <
20 \ps$ are accepted.  

\begin{figure}[tp]
\begin{center}
\epsfig{figure=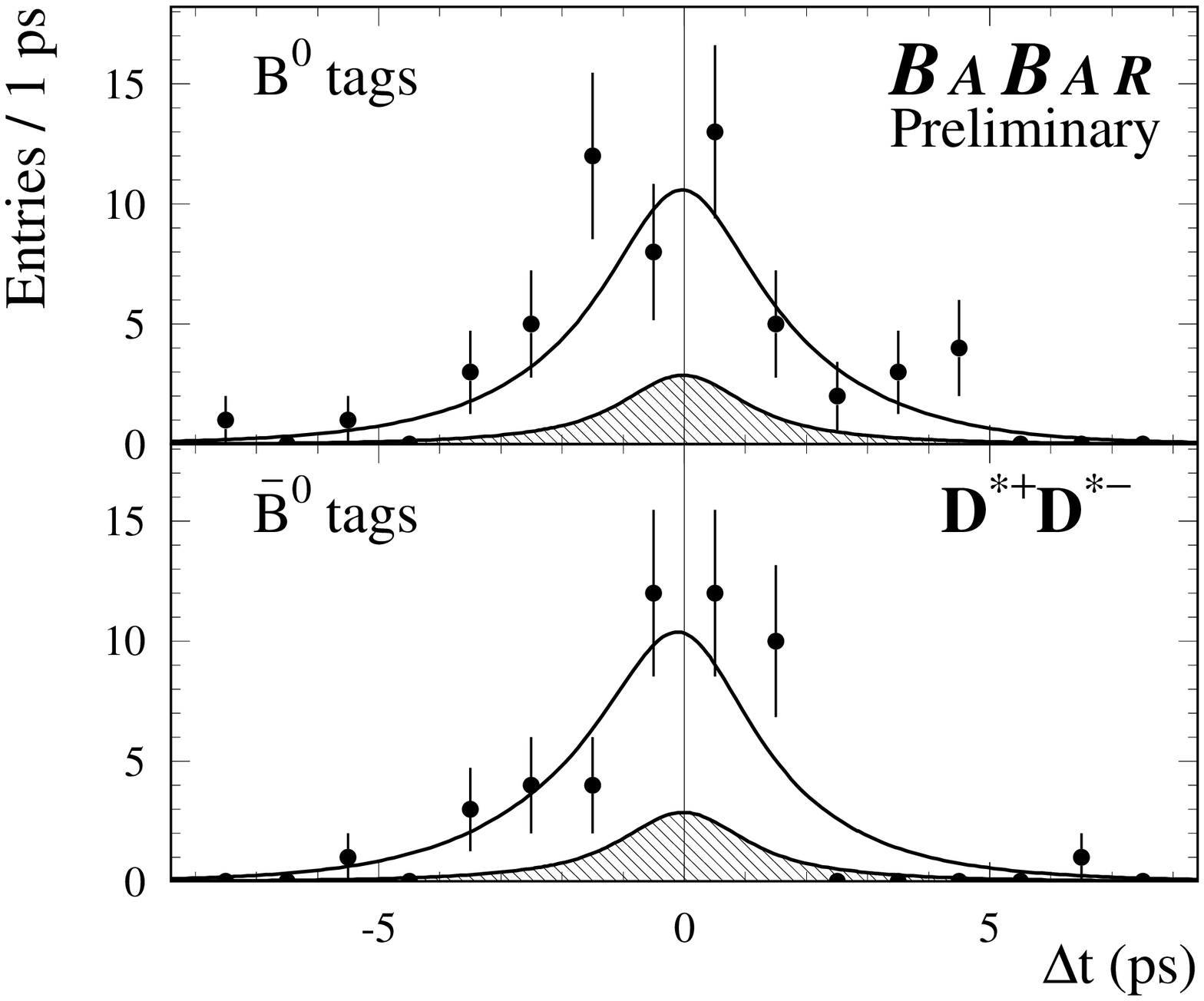,width=0.7\linewidth}
\epsfig{figure=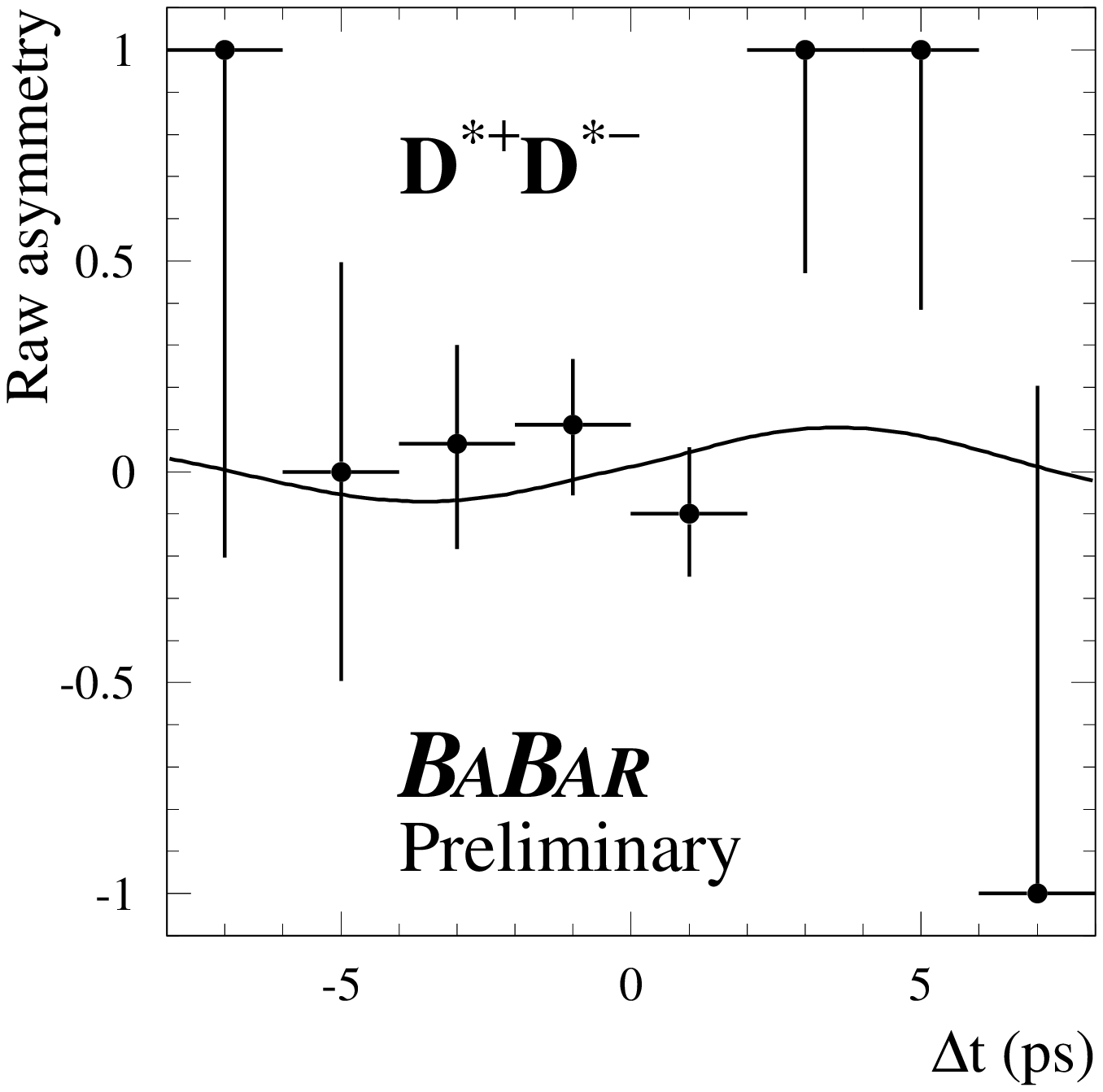,width=0.67\linewidth}\hspace*{2.0cm}
\caption{From top to bottom: number $N_{\Bz }$ of candidate events 
in the signal region with a \Bz tag, number $N_{\Bzb}$ of candidates
with a \Bzb tag, and the raw asymmetry
$(N_{\Bz}-N_{\Bzb})/(N_{\Bz}+N_{\Bzb})$, as functions of \deltat . The
solid curves represent the result of the combined fit 
to the full sample.
The shaded regions represent the background contributions.
}   
\label{fig:cpdeltat}
\end{center}
\end{figure}

We determine the parameters ${\rm Im}(\lambda_{+})$ and $|\lambda_{+}|$ with a simultaneous unbinned maximum likelihood fit 
to the \deltat distributions of the $B_{\rm rec}$ and $B_{\rm flav}$ tagged
samples (Fig.~\ref{fig:cpdeltat}). 
The \deltat distribution of the $B_{\rm flav}$ sample evolves
according to the known frequency for flavor oscillations in neutral $B$
mesons. The observed amplitudes for the \CP asymmetry in the
$B_{\rm rec}$ sample and for flavor oscillation in the $B_{\rm flav}$ sample 
are reduced by the same factor $(1-2\mistag)$ due to flavor mis-tags. The
\deltat distributions for the $B_{\rm rec}$ and $B_{\rm flav}$ samples are
both convolved with a common \deltat resolution function. Events are
assigned signal and background probabilities based on  
their  \mes\ values.
Backgrounds are incorporated with an empirical
description of their \deltat evolution, containing prompt (zero
lifetime) and 
non-prompt components convolved with a separate resolution   
function~\cite{babar-stwob-prd}.

A total of $38$ parameters are varied in the fit, including the values of
${\rm Im}(\lambda_{+})$ and $|\lambda_{+}|$ $(2)$, 
the effective  \CP-odd fraction $(1)$,
the average mis-tag fraction $w$ and the difference $\Delta w$
between $B^{0}$ and $\Bzb$ mis-tags for each tagging
category $(8)$,
parameters for the signal $\Delta t$ resolution $(9)$,
and parameters for the background time dependence $(7)$,
$\Delta t$ resolution $(3)$, and mis-tag fractions $(8)$.
Because the \CP-odd fraction is small, 
the parameters $|\lambda_{\perp}|$ and ${\rm Im}(\lambda_{\perp})$ 
are poorly determined. Therefore they are fixed in the
fit to
$1.0$ and $-0.741$~\cite{babar-stwob-newprl} respectively.  
These are the values expected
if direct \CP
violation and contributions from penguin diagrams are neglected. 
The changes in the fitted values of ${\rm Im}(\lambda_{+})$ and $|\lambda_{+}|$ for 
different input values of ${\rm Im}(\lambda_{\perp})$ (varied between
$-1.0$ and $1.0$) and $|\lambda_{\perp}|$
(varied between 0.7 and 1.3) are taken into account as systematic uncertainties.
The preliminary results obtained from the fit (Fig.~\ref{fig:cpdeltat}) are as follows:
\begin{eqnarray}
{\rm Im}(\lambda_+) & = & 0.31 \pm 0.43(stat) \pm 0.13(syst)\\
|\lambda_+| & = & 0.98 \pm 0.25(stat) \pm 0.09(syst).
\end{eqnarray}

The dominant source of systematic uncertainty comes from the variation on
the value of $\lambda_{\perp}$ (0.09 and 0.02 respectively for ${\rm
  Im}(\lambda_{+})$ and $|\lambda_{+}|$). Other relevant sources are
the angular acceptance and resolution of the detector (0.06 and 0.08),
the level, composition, and \CP asymmetry of the background (0.07 and
0.02), the uncertainty on the SVT internal alignment and boost (0.03 and
0.02), limited Monte Carlo simulation statistics (0.02 and 0.01), 
and possible differences between $B_{\rm flav}$ and  $B_{\rm rec}$
mis-tag fractions and resolution function parameters (0.01 and  0.01).
The total systematic error is 0.13 for ${\rm Im}(\lambda_{+})$ and 
0.09 for $|\lambda_{+}|$.

If the $B \to D^{*+}D^{*-}$ transition proceeds only through the 
$b \to c \bar{c} d $
 tree amplitude, we expect that ${\rm Im}(\lambda_+) = -\sin2\beta$ and
 $|\lambda_+| =1.$
To test this hypothesis, we fix ${\rm
  Im}(\lambda_+)=-0.741$~\cite{babar-stwob-newprl}
 and $|\lambda_+|=1$ 
and repeat the fit. The observed change in the likelihood
corresponds to 2.7 standard deviations (statistical error only). 
More data
is needed to establish whether there are significant contributions from other
processes, in particular, penguin diagrams.

\section{Summary}
\label{sec:Summary}
We have reported preliminary measurements of time-dependent \CP
asymmetries and a measurement of the \CP-odd fraction for the decay
\Bztodstdst.  The measurement of $R_\perp$ represents 
an improvement in the statistical uncertainty by a factor of almost three
compared to previous measurements.
The time-dependent asymmetry
measurements still have large statistical uncertainties. 
These should be reduced steadily in the coming years as \babar\ 
accumulates
additional data, thus allowing useful tests of the
Standard Model.

\section{Acknowledgments}
\label{sec:Acknowledgments}
We are grateful for the 
extraordinary contributions of our \pep2\ colleagues in
achieving the excellent luminosity and machine conditions
that have made this work possible.
The success of this project also relies critically on the 
expertise and dedication of the computing organizations that 
support \babar.
The collaborating institutions wish to thank 
SLAC for its support and the kind hospitality extended to them. 
This work is supported by the
US Department of Energy
and National Science Foundation, the
Natural Sciences and Engineering Research Council (Canada),
Institute of High Energy Physics (China), the
Commissariat \`a l'Energie Atomique and
Institut National de Physique Nucl\'eaire et de Physique des Particules
(France), the
Bundesministerium f\"ur Bildung und Forschung and
Deutsche Forschungsgemeinschaft
(Germany), the
Istituto Nazionale di Fisica Nucleare (Italy),
the Research Council of Norway, the
Ministry of Science and Technology of the Russian Federation, and the
Particle Physics and Astronomy Research Council (United Kingdom). 
Individuals have received support from 
the A. P. Sloan Foundation, 
the Research Corporation,
and the Alexander von Humboldt Foundation.


\begin{thebibliography}{99}

\bibitem{babarCP}
B.~Aubert, {\it et al.}, \jprl {\bf 87}, 091801 (2001).

\bibitem{belleCP}
K.~Abe, {\it et al.}, \jprl {\bf 87}, 091802 (2001).

\bibitem{CKM}
N.~Cabibbo, \jprl {\bf 10}, 531 (1963);\\
M.~Kobayashi and T.~Maskawa, \progtp {\bf 49}, 652 (1973).

\bibitem{aleksan}
R.~Aleksan, {\it et al.}, \plb {\bf 317}, 173 (1993).

\bibitem{sanda}
A.I.~Sanda and Z.Z.~Xing, \jprd {\bf 56}, 341 (1997).

\bibitem{xing}
X.Y.~Pham and Z.Z.~Xing, Phys. Rev. B {\bf 458}, 375 (1999).

\bibitem{babar-stwob-newprl}
\babar\ Collaboration, B.\ Aubert {\it et al.}, SLAC-PUB-9293,
hep-ex/0207042, submitted to \jprl

\bibitem{xing2}
Z.Z.~Xing, \jprd {\bf 61}, 014010 (2000).

\bibitem{angular} I.~Dunietz, {\it et al.}, \jprd {\bf 43}, 2193 (1991). 

\bibitem{ref:dstdstprl} B.~Aubert, {\it et al.}, SLAC-PUB-9152, \hepex{0203008}, to appear in \jprl

\bibitem{ref:babar}
The \babar\ Collaboration, A.\ Palano {\em et al.},
Nucl.\ Intrum.\ Methods. {\bf A479}, 1 (2002).

\bibitem{ref:fox}
G.~C.~Fox and S.~Wolfram, \jprl {\bf 41}, 1581 (1978).

\bibitem{pdg} Particle Data Group, K.~Hagiwara {\it et al.},
\jprd{66}, 010001 (2002).

\bibitem{argus} Defined as $A \sim \sqrt{1- (m_{ES}/m_0)^2} \times
\exp({\kappa}(1- (m_{ES}/m_0)^2)$ for $m_{ES} < m_0$.  The
value of $m_0$ is fixed to 5.291\,\gevcc.
ARGUS Collaboration, H.~Albrecht {\it et al.}, \zpc {\bf 48}, 543 (1990).

\bibitem{penguin} P.~F.~Harrison and H.~R.~Quinn, eds. ``The \babar\
Physics Book'', SLAC-R504 (1998), Chapter 5, and references therein.

\bibitem{babar-stwob-prd}
\babar\ Collaboration, B.\ Aubert {\it et al.}, SLAC-PUB-9060,
hep-ex/0201020, to appear in \jprd.

\end{thebibliography}
\end{document}